\documentclass[conference]{IEEEtran}
\IEEEoverridecommandlockouts
\usepackage{cite}
\usepackage{amsmath,amssymb,amsfonts}
\usepackage{algorithmic}
\usepackage{graphicx}
\usepackage{textcomp}
\def\BibTeX{{\rm B\kern-.05em{\sc i\kern-.025em b}\kern-.08em
    T\kern-.1667em\lower.7ex\hbox{E}\kern-.125emX}}
\begin{document}

\title{Intent Classification using Feature Sets for
	Domestic Violence Discourse on Social Media
}

\author{\IEEEauthorblockN{Sudha Subramani}
\IEEEauthorblockA{\textit{Centre for Applied Informatics} \\
\textit{Victoria University}\\
Melbourne, Australia \\
sudha.subramani1@live.vu.edu.au
}
\and
\IEEEauthorblockN{Huy Quan Vu}
\IEEEauthorblockA{\textit{Centre for Applied Informatics} \\
\textit{Victoria University}\\
Melbourne, Australia \\
huyQuan.vu@vu.edu.au
}
 \and
\IEEEauthorblockN{Hua Wang}
\IEEEauthorblockA{\textit{Centre for Applied Informatics} \\
\textit{Victoria University}\\
Melbourne, Australia \\
hua.wang@vu.edu.au}
}
\maketitle

\begin{abstract}
Domestic Violence against women is now recognized to be a serious and widespread problem worldwide. Domestic Violence and Abuse is at the root of so many issues in society and considered as the societal tabooed topic.  Fortunately, with the popularity of social media, social welfare communities and victim support groups facilitate the victims to share their abusive stories and allow others to give advice and help victims. Hence, in order to offer the immediate resources for those needs, the specific messages from the victims need to be alarmed from other messages. In this paper, we regard intention mining as a binary classification problem (abuse or advice) with the use-case of abuse discourse. To address this problem, we extract rich feature sets from the raw corpus, using psycholinguistic clues and textual features by term-class interaction method. Machine learning algorithms are used to predict the accuracy of the classifiers between two different feature sets. Our experimental results with high classification accuracy give a promising solution to understand a “big social problem” through big social media and its use in serving information needs of various community welfare organizations.
\end{abstract}


\section{Introduction}
Domestic Violence (DV) is a global issue of pandemic proportions and vulnerable to any age group, culture, socioeconomic group, and education level. The World Health Organization estimates that 35\% of women worldwide have experienced Intimate Partner Violence (IPV)\cite{garcia2013global}. IPV refers not only to physical, but it also includes sexual, verbal, psychological and financial \cite{world2012understanding}. DV has severe and persistent effects on physical health and also have a cumulative impact on women mental health \cite{bromfield2010issues}. Hence, there is a burning need to better characterize and understand DV, to provide appropriate resources for victims and efficiently implement control measures.

People are increasingly using social media platforms, such as Twitter and Facebook. They share their thoughts and opinions of daily activities and happenings on these sites in a naturalistic setting. Thus generates massive amount of user data discussing various topics, even the socially stigmatized context and society tabooed topics like DV. Victims experiencing abuse are in need of earlier access to specialized DV services such as health care, crisis support, legal guidance and so on. They may not have the option or unaware to seek help directly. In social media, many social welfare and non profit organizations are encouraging the victims and survivors of DV to express their feelings and share their experience of being in an abusive relationship. Hence the social support groups for a good social cause play a leading role in creating awareness promotion and leveraging various dimensions of social support like emotional, instrumental, and informational support to the victims. When the victims seek help, it is important for those groups to identify those critical posts and provide a clear call-to-response help with more immediate impact. From the citizen-generated data, social welfare organizations with limited resources are trying to incorporate information nuggets to enrich their decision support system \cite{purohit2015intent}.

Intent mining provides insights that are not explicitly available from the user generated data. Intent is defined as a purposeful action and this can help  to identify actionable information \cite{varga2013aid} \cite{purohit2013emergency}. Intent classification (focused on future action) is a form of text classification. In our work, we can apply intent classification to classify the user posts into any of the two classes as abuse or advice. If the user shared their life experience of abusive relationship, that post is classified in ``abuse" class. Instead, if the post relates to awareness promotion, giving advice or  opinion, it is to be classified in ``advice/opinion" class. 

\begin{table*}[]
	\centering
	\caption{Example posts and its associated intent class}
	\label{tab1:ex_posts}
	\resizebox{\linewidth}{!}{  	% copy this line
	\begin{tabular}{|c|l|l|}
		\hline
		\textbf{S.No} & \multicolumn{1}{c|}{\textbf{POST}}                                                                                                                                                                                                                                                                                                                    & \multicolumn{1}{c|}{\textbf{CLASS}} \\ \hline
		1             & To understand why people stay in abusive relationships. Visit the link.                                                                                                                                                                                                                                                                               & advice/opinion                      \\ \hline
		2             & \begin{tabular}[c]{@{}l@{}}If I could go back to the day that I was bickering with my first husband. \\ Nothing serious just small disagreement.\\ I was 8 months pregnant walked in the kitchen to get a drink and boom he was hiding and waiting. \\ It took one punch to my head I fell backwards out unconscious in a pool of blood.\end{tabular} & abuse                               \\ \hline
		3             & \begin{tabular}[c]{@{}l@{}}Negative emotions like hatred destroy our peace of mind. \\ Guide your- self to find peace and healing within you and other people. \\ Hope we become better people to love and share the best in us .Bless.\end{tabular}                                                                                                  & advice/opinion                      \\ \hline
		4             & \begin{tabular}[c]{@{}l@{}}Yesterday I lost a loving and dear sweet friend. \\ Her life was cut short by the monster she once Loved and called her husband.\end{tabular}                                                                                                                                                                              & abuse                               \\ \hline
	\end{tabular}
}
\end{table*}

From the table \ref{tab1:ex_posts} , we can clearly understand that the posts 2 and 4 describe the story about abusive relationship. The post 2 is shared by DV survivor and post 4 is shared by victi{m'}s friend. The post 1 has some awareness promotion in it and post 3 expresses on{e}'s thought/advice. Hence, they are classified as opinion/advice. Thus, our research question is ``{How to mine relevant social intent from an ambiguous and proliferation of unstructured textual data in abuse discourse?}" By notion, the relevant intent classes meet actionable information needs of an organization in a given context \cite{purohit2015intent}. There is no chance of helping the victim or victim's family, if the message is unnoticed or ignored. Thus the paper focuses to identify abuse related posts and turn that knowledge into action that will help the victims of abuse. Thus the intent classification system in turn provide an actionable knowledge, which helps the society in turn.

With the proliferation of unstructured data, text classification or text categorization has found many applications in topic classification, sentiment analysis, authorship identification, spam detection, and so on. We observed two key challenges with intent classification on our DV discourse. First, informal language use in short-text messages creates ambiguity to interpret user expressions and thus weakening term-class relationships. Second, sparsity of instances of specific intent classes in the corpus creates data imbalance. In the binary classification, both intent classes may co-occur within a single message. For instance, post 1 in table \ref{tab1:ex_posts}  may be classified into class abuse instead of class advice, as the text contains keyword abuse. 
Hence in our work, intent classification exploits a rich feature representation for learning, created by using knowledge sources from psycholinguistics features, and also the textual features from the underlying post. We extract the texts relating to DV from Facebook, because it allows for long text discussion and therefore use of standard English is more common.Facebook allows users to comment on posts, providing them with the ability to share their stories about abusive life pattern, give advice, and provide support. The two contributions of this work are as follows:
\begin{itemize}

\item Constructing the two different and efficient feature sets by analysing the psycholinguistic dimensions and the textual features of the user postings on Facebook.
\item Evaluating the classifiers performance for identifying texts by constructing and comparing the two different feature set from DV discourse.

\end{itemize}

Analysis of the linguistic structures embedded in these posts instances provides insight the victims of domestic abuse report their personal story and they need emergency support. Trained classifiers agree with these linguistic structures, adding evidence that these social media texts provide valuable insights into DV. Intent mining classification can help to design efficient cooperative information systems between citizens and organizations for serving organizational information needs and help the victims in need.
  
The structure of the paper is as follows. In Section \ref{sec:rel_work}, the related work is discussed. Section \ref{sec:prb_stm} defines the problem statement and experimental analysis is discussed in Section \ref{sec:exp_ana}. Section  \ref{sec:feature}  discusses feature extraction method in detail and Section \ref{sec:classifi} explains about classifiers and evaluation metrics. Prediction results are discussed in Section \ref{sec:predict}. Finally, Section \ref{sec:concl} discusses about the conclusion.

\section{Related work} \label{sec:rel_work}

With the increasing popularity of social media, the amount of information now available to decisive moment is massive. The sheer overwhelm of social media data makes it to be the one of Big Dat{a}'s most significant sources \cite{wang2015special} \cite{qin2016things}. Several research studies focused on social media to analyse and predict real world activities like user sentiment analysis, opinion mining on political campaigns \cite{tumasjan2010predicting} \cite{o2010tweets}, natural disasters \cite{sakaki2010earthquake}, epidemic surveillance \cite{chunara2012social}, event detection \cite{petrovic2010streaming}, tourism \cite{vu2015exploring} \cite{vu2017travel}, e-healthcare services \cite{sun2012purpose}\cite{sun2012semantic} and so on.  On the other hand, some studies dealt with security and privacy issues \cite{wang2017special} \cite{sun2012satisfying} \cite{li2008current} \cite{wang2005flexible} \cite{wang2010trust}, as the increasing sophistication of social media data is posing substantial threats to users privacy.  In contrast to traditional media, social media becomes a fast reaching data source to share the opinions and thoughts in online immediately with a status update. Hence, this becomes an efficient platform for researchers to detect real world events in informational retrieval and decision making. 

Recent studies have predicted mental health conditions by analysing textual features \cite{nguyen2015autism}\cite{nguyen2014affective}\cite{nguyen2017using}. For this kind of prediction, two main characteristics such as topics and language styles expressed via text have been popularly investigated. For identifying depression, linguistic styles such as an expression of sadness or the use of swear words have been used as the cues \cite{rodriguez2010reading}.

Linguistic Inquiry and Word Count (LIWC) \cite{pennebaker2015development} has been commonly used to capture language characteristics and they are considered to be influential predictors of depression-related disorders and mental health \cite{ramirez2008psychology} \cite{stirman2001word}. Popular Bayesian probabilistic modelling tools, such as Latent Dirichlet allocation (LDA) are used to extract the topics \cite{huang2017probabilistic}. LDA and its variants have been used previously to discover several mental ailments discussed in millions of tweets \cite{paul2014discovering}. The authors \cite{balani2015detecting} used standard n-gram features,  submission length and author attributes to classify a Reddit submission as high or low level self-disclosure. Another study states that, support seeking in stigmatized context of sexual abuse on Reddit and investigated the use of throw away accounts in the aspects of support seeking, anonymity and first time disclosures \cite{andalibi2016understanding}. Most relevantly, Schrading et al. \cite{schrading2015whyistayed} analyzed the lexical structures of the tweet to predict whether a victim was about staying or leaving an abusive relationship.

Though, the majority of studies assessing the role of Twitter, Facebook, and Reddit for the enormous new findings, there is limited research that focuses on online DV disclosures. Hence, this is the first study to conduct an intent classification of Facebook content that relates to abuse discourse. 

\section{Problem Statement} \label{sec:prb_stm}

 Let  $ B=\begin{Bmatrix} m_i|1\leqslant i\leqslant n
 \end{Bmatrix}  $  denote the corpus of all $n$ Facebook posts, each post $ m_i $ is a facebook post generated in DV community pages,  and  $ L=\begin{Bmatrix}
		L_j|j=<abuse>,<advice/opinion>
	\end{Bmatrix} $ is a set of binary classes.
	
 For each given post $m_i\epsilon B$, predict an intent class in $L$ based on textual features $f_j$ extracted from post:
	$F^{(m_i)}=[....,f_j^{(m_i)}.......|1\leqslant j\leqslant p]$.

	To characterize the difference between two classes of DV dataset, two feature sets are extracted.\\

\begin{figure}
	\centering
	\includegraphics[width=1.0\linewidth]{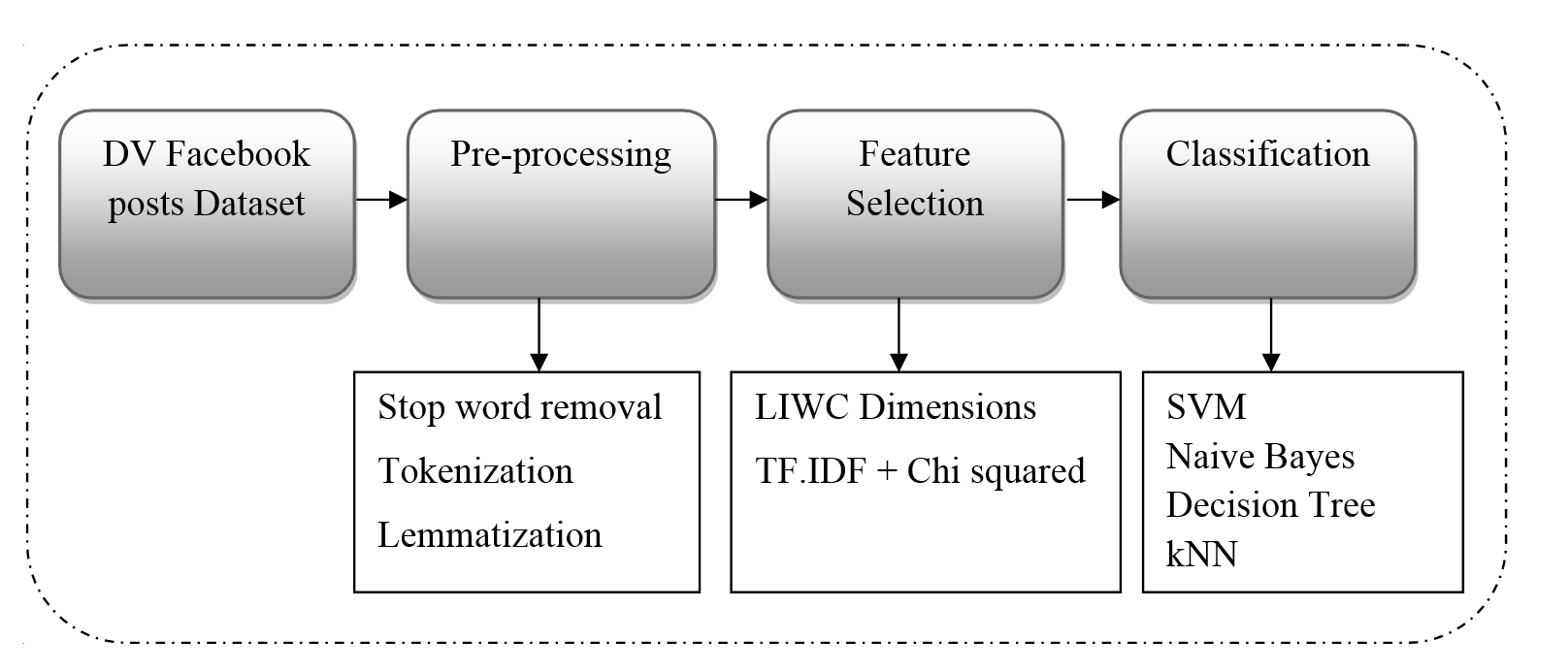}
	\caption{Architecture of automated classification based on training feature sets}
	\label{fig:arch2}
\end{figure}

\begin{itemize}

 \item LIWC Features: Psycholinguistic knowledge based on LIWC [16] are used as feature sets. These features differentiate the semantic-syntactic patterns and informational context of two different classes of abuse and advice. Thus training the classifiers for intent classification and improve the accuracy.\\

 \item Term based model: Feature sets are built based on textual features derived from Term-Class interaction with chi-square metric [24] to synthesize a more accurate classification procedure.\\

\end{itemize}
The model for the current study is illustrated in Fig. ~\ref{fig:arch2} and in the following sections, we describe the components of this model.

\section{Experimental Analysis} \label{sec:exp_ana}

Effective intent mining on social media data is demanding due to ambiguity in interpretation of the textual data, and its sparsity of relevant behaviour \cite{purohit2015intent}. In this paper, we address the binary classification of intent with a use-case of DV data generated in Facebook. This study uses machine learning approaches to discriminate the online posts into two categories based on two different feature set approaches.  The first approach which is explained in section \ref{sec:method1} exploits the use of psycholinguistic features generated with LIWC \cite{pennebaker2015development} and has the advantage to achieve higher accuracy and also being computationally efficient. Another approach explained in section \ref{sec:method2} constructs the feature set based on bag-of-tokens \cite{sparck1972statistical}, which uses tf.idf  approach combined with chi-squared algorithm \cite{legendre2012numerical}. Thus, the classifier is finally trained with higher accuracy based on the constructed feature set. 

\subsection{Data Collection} \label{sec:data_coll}

For the current study, we extracted the data from Facebook pages which is focusing on specific theme like DV using NCapture\cite{Ncapture}. The dataset contains 8856 posts and 28873 comments and extracted posts are ranging from historical to current date (2014 to 2017). The data extracted contains information about post content, user name, link description, comment text, commenter username, created time, and likes. Table \ref{tab2:data collect} illustrates our dataset details. 
\begin{table}[]
	\centering
	\caption{Details of DV Facebook page name, posts, comments and likes*(till date*)}
	\label{tab2:data collect}
	\begin{tabular}{|l|l|l|l|}
		\hline
		\multicolumn{1}{|c|}{\textbf{DV Page Source Name}} & \multicolumn{1}{c|}{\textbf{Posts}} & \multicolumn{1}{c|}{\textbf{Comments}} & \multicolumn{1}{c|}{\textbf{Page Likes}} \\ \hline
		Stop DV                                            & 3597                                & 13462                                  & 63k                                      \\ \hline
		DV Survivors                                       & 2816                                & 8415                                   & 32k                                      \\ \hline
		Stop DV against Women                              & 516                                 & 2802                                   & 7.5k                                     \\ \hline
		DV Awareness Month                                 & 1927                                & 4194                                   & 15k                                      \\ \hline
	\end{tabular}
\end{table}

\subsection{Extracting Gold Standard Labels} \label{sec:gold}

Two of the authors annotated a random sample of 1135 posts to classify the corpus and labelled them as abuse and opinion/advice based on textual information. Hence 510 posts are classified in first class, i.e 45\% and 625 posts are labelled in second class which is 55\%. The kapp{a}'s coefficient was good between the authors and it was 0.85. To measure the ``inter-rater reliability",  a statistical measure Kappa coefficient is used to measure the degree of agreement between two users. Any discrepancies between the authors are clarified after further discussion.

\subsection{Data Pre-processing} \label{sec:prepro}

For the effective analysis, text pre-processing is the most important step, as it removes noise that produces negatives effects and degrades performance.  High quality information and features are extracted by incorporating some pre-processing techniques like stop word removal and some normalization techniques like stemming, lemmatization and so on. First, we removed stop words that are not included for the content analysis.  Stop word lists contain common English words like articles, prepositions, conjunctions, pronouns, etc., Few examples are the, a, an, the, in, and at.  Next, pre-processing step is lemmatization. This is used to reduce more inflectional forms of words into a more limited form of canonical forms. This helps to standardize terms appearance and to reduce data sparseness. For example, the following terms such as physically assaulted, physically assaults and physically assaulting are all lemmatized to ``physical assault". Similarly physically abusing, physically abused, physically abuses are all lemmatized to ``physical abuse".  

\section{feature Extraction} \label{sec:feature}

Feature Extraction is one of the most important steps in data mining and knowledge discovery. The idea is to select the best features $f_j$ that improve the classification accuracy. In the following sections \ref{sec:method1} and \ref{sec:method2}, the two different feature extraction techniques used in this work is discussed. 

\subsection{Methodology 1: Psycholinguistic Features Analysis} \label{sec:method1}

We examine and analysed the proportions of word usage in psycholinguistic categories as defined in the LIWC 2015 package \cite{pennebaker2015development}. The LIWC analyses text on a word-by-word basis and calculates the percentages of words that match particular word categories.

For each given post  $m_i\epsilon B$ corpus, predict an intent class in $L$ based on LIWC features extracted from post $m_i$ defined as  
$F^{(m_i)}=[....,f_j^{(m_i)}.......|1\leqslant j\leqslant p] $. Here $f_j^{(m_i)}$ denotes the quantity of specific psycholinguistic feature $j$
in post $m_i\epsilon B$

LIWC package is a psycholinguistic lexicon created by psychologists with focus on identifying the various emotional, cognitive, and linguistic components present in individual{s}’ verbal or written communication. For each input of a post, it returns more than 70 output variables with higher level hierarchy of psycholinguistic features such as 
\begin{itemize}
\item linguistic dimensions, other grammar, informal language 	
\item temporal, affective, social processes 
\item cognitive, biological,  perceptual  processes
\item personal concerns, drives, relativity.
\end{itemize}
 These higher level categories are further specialized in sub-categories such as in 
\begin{itemize}
\item biological processes - body, sexual, health and ingestion.
\item affective processes - positive emotion, negative emotion and negative emotion further sub-classified as anger, anxiety, and sadness.
\item drives - affiliation, acheivement, power, regard, and risk.
\end{itemize}

For evaluating the prediction accuracy of psycholinguistic features, each individual facebook post is converted to a vector of 70 output numerical variables, as mentioned above. Each output variable represents the frequency distribution of the appearance of those categories appeared in the specific post. Each word in the post could fit some categories and not fit into some categories. Hence, there would be the huge difference between the posts, to which category it belong. For instance, the following  post ``\textit{Please view, share and is possible donate.  We appreciate your support!} " has higher value of `positive emotion' (36.36\%), `focus present' (36.36\%), `you' (9.09\%), `social' (27.27\%) and has (0\%) for the categories such as `negative emotion', `shehe', `bio', `body'. The above post falls into ``advice/opinion" category, as it creates a good social cause and for fund-raising, and also has higher percentage of positive expression and present focus in it. In contrast, the post ``\textit{He is just an evil, greedy, arrogant little man}" has higher percentage of `negative emotion'(40\%), `{anger}' (10\%) , `male'(20\%), `shehe'(10\%) and (0\%) of `posemo', `you', `death'. This post falls into ``abuse'' class, as woman explaining about her abusive partner and the post carries a lot of negative emotion in it.

\subsubsection{Most informative features and analysis} We only selected the top-most informative 15 features,  as shown in table \ref{tab3:liwc} to perform the binary classification task. We selected those features based on the mean value of all the posts, which is defined in table \ref{tab4:mean_score}.  We selected the feature sets that are strong predictors of two classes such as ``advice/opinion" and ``abuse". 

LIWC sub-categories such as `negative emotion', `anger', `shehe', `focuspast' are  features with higher mean value and good prediction level for ``abuse" category. `Positive emotion', `focus present', `you', `focus future' sub-categories, as expected, are good predictors for ``opinion/advice" class.  Another important prediction is that `health', `sexual issues', and personal concern such as `death' are as good predictors for ``abuse" class. The results infer that, because of the abusive cycle, most of the victims suffer from severe health issues and also death. Further analysis show that posts related to ``abuse" category are often self-reflective, with more words related to personal pronouns i.e, usage of more pronouns such as `I' and `shehe', when describing their life experience about violence, whereas in the ``opinion/advice" category, $2^{nd}$ person usage `you' is higher, when giving advice or sharing opinion to other people. It is important to compare the time orientations, the posts of ``abuse" category are more focused on `past' and contains negative emotions with expression of `anxiety', `angry' and `sad'.  On the other hand, the ``advice" category contains more of positive emotion, as sharing of good thoughts and opinions and more time orientated towards `present' and `future'. 

\begin{table}[]
	\centering
	\caption{LIWC features and the example words used in our dataset}
	\label{tab3:liwc}
	\resizebox{\linewidth}{!}{
	\begin{tabular}{|l|l|l|}
		\hline
		\multicolumn{1}{|c|}{\textbf{Category}} & \multicolumn{1}{c|}{\textbf{Dimension}}                                         & \multicolumn{1}{c|}{\textbf{Example words}}                                                                                                             \\ \hline
		Linguistic Dimensions                   & \begin{tabular}[c]{@{}l@{}}personal\\ pronouns(I,you,shehe)\end{tabular}        & I, you, he,she ,his, him, her,herself                                                                                                                   \\ \hline
		Time orientations                       & \begin{tabular}[c]{@{}l@{}}focuspast\\ focuspresent\\ focusfuture\end{tabular}  & \begin{tabular}[c]{@{}l@{}}broke,ran, accepted\\ supports,trust, likes\\ plan,wish, hopeful\end{tabular}                                                \\ \hline
		Biological Processes                    & \begin{tabular}[c]{@{}l@{}}body\\ sexual\\ health\end{tabular}                  & \begin{tabular}[c]{@{}l@{}}muscles, injury, fat\\ rape, lust, abortion, pregnant\\ sick, weak, painful, bleed\end{tabular}                              \\ \hline
		Psychological Processes                 & \begin{tabular}[c]{@{}l@{}}posemo\\ negemo\\ anxiety\\ anger\\ sad\end{tabular} & \begin{tabular}[c]{@{}l@{}}hope,share,support,like\\ threat,lose,hate\\ threat,misery,worry\\ sucks,hate,yell\\ miss,lose,suffer,overwhelm\end{tabular} \\ \hline
		Personal Concern                        & death                                                                           & die, murder, kill, suicide, bury                                                                                                                        \\ \hline
	\end{tabular}
}
\end{table}

The posts related to ``abuse" category has higher mean values of the corresponding features such as `I', `shehe', `focuspast', `body', `sexual', `health', `death' and `negative emotions'. The posts related to "opinion/advice" scores high in features such as you, focuspresent, focusfuture, posemo. For eg., if we consider  Shehe feature, the posts belongs to "abuse" category have the highest mean value of 10.98, whereas the posts of ``opinion/advice" category have the lowest mean value of 0.39. This implies, when the victims post their story, they need to say more about the abuser and thus used more third person pronoun `shehe'. In the ''opinion/advice" class, the people just express their thoughts and not necessarily use third person pronoun.

\begin{table}[]
	\centering
	\caption{ Mean Scores of Psycholinguistic Processes (LIWC) for the posts of 2 categories}
	\label{tab4:mean_score}
		\resizebox{\linewidth}{!}{
	\begin{tabular}{|l|l|l|l|l|l|}
		\hline
		\textbf{Features} & \textbf{Abuse ($\mu$)} & \textbf{Advice($\mu$)} & \textbf{Features} & \textbf{Abuse ($\mu$)} & \textbf{Advice($\mu$)} \\ \hline
		I                 & 3.15               & 2.02               & Health            & 1.01               & 0.61               \\ \hline
		You               & 0.59               & 4.01               & Death             & 1.19               & 0.02               \\ \hline
		Shehe             & 10.98              & 0.39               & Posemo            & 2.57               & 10.73              \\ \hline
		Focuspast         & 7.09               & 0.10               & Negemo            & 4.78               & 1.55               \\ \hline
		Focuspresent      & 7.11               & 14.34              & Anxiety           & 0.65               & 0.09               \\ \hline
		Focusfuture       & 0.96               & 1.55               & Anger             & 2.28               & 0.57               \\ \hline
		Body              & 0.84               & 0.42               & Sad               & 0.58               & 0.14               \\ \hline
		Sexual            & 0.34               & 0.09               &                   &                    &                    \\ \hline
	\end{tabular}
}
\end{table}

\subsection{Methodology 2:  tf.idf approach with chi-squared metric 
	($\chi^2$  Statistic) as feature selection parameter} \label{sec:method2}

\begin{itemize}
	
	\item In Term Document Matrix  $TM(D*T)$ dimension, $D$ indicates the total number of posts and $T$  indicates no. of terms. Thus $TM_{ij}$  indicates the corresponding $tf.idf$ matrix. Two common features such as  $TF$ and $IDF$ used, where $TF(t,p)$  represents the number of times the term $t$ appears in post $p$ and $DF(t)$ denotes the number of posts contain term $t$. 
	
	$TF.IDF(t,p)$ weighting scheme improves the discriminative power, where $TF.IDF(t,p) = TF(t,p)× IDF(t)$ with $IDF(t) = |B|/DF(t)$ is the inverse document frequency. In this work, a term t will be selected if it has high $DF(t)$ value and high average values of  $TF(t,p)$ and thus high $TF.IDF(t,p)$ across all $D$ Facebook posts over a threshold. Term-Class interaction based selection method is also used which capture the dependence between terms and corresponding class labels during the feature selection process.
	
	\item The feature selection technique based on chi-squared on the entire term document matrix $TM(D*T)$ is used to compute chi-squared value corresponding to each word. 
	
	The Chi-square statistical test has been widely accepted as a statistical hypothesis test to evaluate the dependency among two variables [24]. In natural language processing, Chi-squared is generally used to measure the degree of dependence between term $t$ and label $l$ and compared to the distribution with one degree of freedom. The expression for ($\chi^2$  Statistic) is defined as
	
	\begin{equation}
	\label{EQ:5}
	\resizebox{0.9\linewidth}{!} 
	{
		$\chi^2_{(t,l)}=\frac{D*(PN-MQ)^2}{(P+M)*(Q+N)*(P+Q)*(M+N))}$
	}
	\end{equation}
	$ D $= the total number of posts. \\
	$P $ = the number of posts of label l containing term t.\\
	$ Q  $= number of posts containing t occurring without l.\\
	$ M  $= number of label l occurring without t.\\
	$ N $ = number of posts of other classes without t. \\
	
\end{itemize}

\section{Classifiers and Evaluation Metrics} \label{sec:classifi}

We now pursue the use of supervised learning to construct classifiers trained to predict the class label of the posts. Although we analysed results for both all dimension-inclusive and dimension-reduced cases, we employ principal component analysis (PCA) to avoid over-fitting for all the classifiers. We compare several different classifiers such as Support Vector Machine, k-Nearest Neighbor, Naive bayes, and Decision Tree to empirically determine the best suitable classification technique. The problem of binary text classification problem is generally defined as follows:

Our training set corpus $B= {m_1,m_2,....,m_n}$ of $n$ posts, such that each post $m_i$ is labelled with a class of either $L={l_1|l_2}$. The task of a classifier $f$ is to find the corresponding label for each post.
\begin{equation}\label{eq:clas}
f : B \epsilon L 	f(m)= l
\end{equation}

For all of our analyses, we use 10-fold cross validation and leave one out methods to assess the effectiveness of the model.\\

\begin{itemize}
	\item Support Vector Machines (SVM): a non-probabilistic generalized and linear binary classifier \cite{cortes1995support}.This maps an input feature vector into a higher dimensional space and find a hyperplane that separates the data into two classes with the maximal margin between the closest samples in each class. \\
	
	\item  Naive Bayes (NB): a probabilistic method \cite{nigam2000text} for text classification is familiar  for its robustness and relative simplicity. This classifier constructs the conditional probability distributions of underlying features given a class label from the training  data only. The classification on unseen data is then performed by comparing the class likelihoods \cite{wang2016detection} \cite{yi2009privacy}.\\	
	
	\item Decision Tree (DT): an interpretable classifier \cite{quinlan1986induction} creates the hierarchical tree of the training instances, in which a condition on the feature value is used to divide the data hierarchically. For the classification of text documents, the conditions on decision tree nodes
	are commonly defined in terms and a node may be subdivided to its children based on the presence or absence of a particular term in the document. Ensemble methods use multiple learning algorithms of decision tree for better predictive performance \cite{hu2006maximally} \cite{wang2015multi} \cite{subramani2016mining}. \\

	\item k-Nearest Neighbor (k-NN): a proximity-based classifier \cite{han2001text} use distance-based measures i.e., the documents which belong to the same class are more  likely similar or close to each other based on the similarity measures. The classification of the test document is reported from the class labels of the k nearest similar documents in the training set.\\
			
\end{itemize}
	
	We have performed various runs with different feature set in the above defined classifiers. We used the common metrics such as Precision, Recall, F-Measure, and Accuracy to evaluate the classification performance. Precision measures the percentage of the Facebook posts that the classifier predicted (i.e., the classifier labeled as positive) that are in fact positive (i.e., are positive according to the human gold labels). Recall measures the percentage of posts actually present in the gold label that were correctly identified by the Classifier. F-measure comes from a weighted harmonic mean of precision and recall. Accuracy calculates the percentage of correctly classified posts versus. Number of total posts. All the metrics are defined as follows.

	\begin{equation}\label{eq:1}
	Precision\left ( P \right ) =\frac{True Positive}{True Positive+ False Positive} 
	\end{equation}
	
	\begin{equation}\label{eq:2}
	Recall\left ( R \right ) =\frac{True Positive}{True Positive+ False Negative} 
	\end{equation} 
	
	\begin{equation}\label{eq:3}
	F-Measure = 2 \frac {PR}{P+R} 
	\end{equation}
	
	\begin{equation}\label{eq:4}
	\resizebox{\linewidth}{!} 
	{
		$ Accuracy=\frac{True Positive+True Negative}{True Positive+True Negative+False Positive+False Negative} $
	}
	\end{equation}

\section{Prediction Results and Discussion} \label{sec:predict}

\subsection{LIWC analysis}

Our chosen most informative features of LIWC have higher accuracy in prediction of two different classes abuse and advice. Among all the classifiers, SVM outperforms all the other classifiers. The classification accuracy of SVM, kNN and decision tree are 97\%, 95.3\%, 95.1\% respectively. 
 
\begin{figure}
	\centering
	\includegraphics[width=1.0\linewidth]{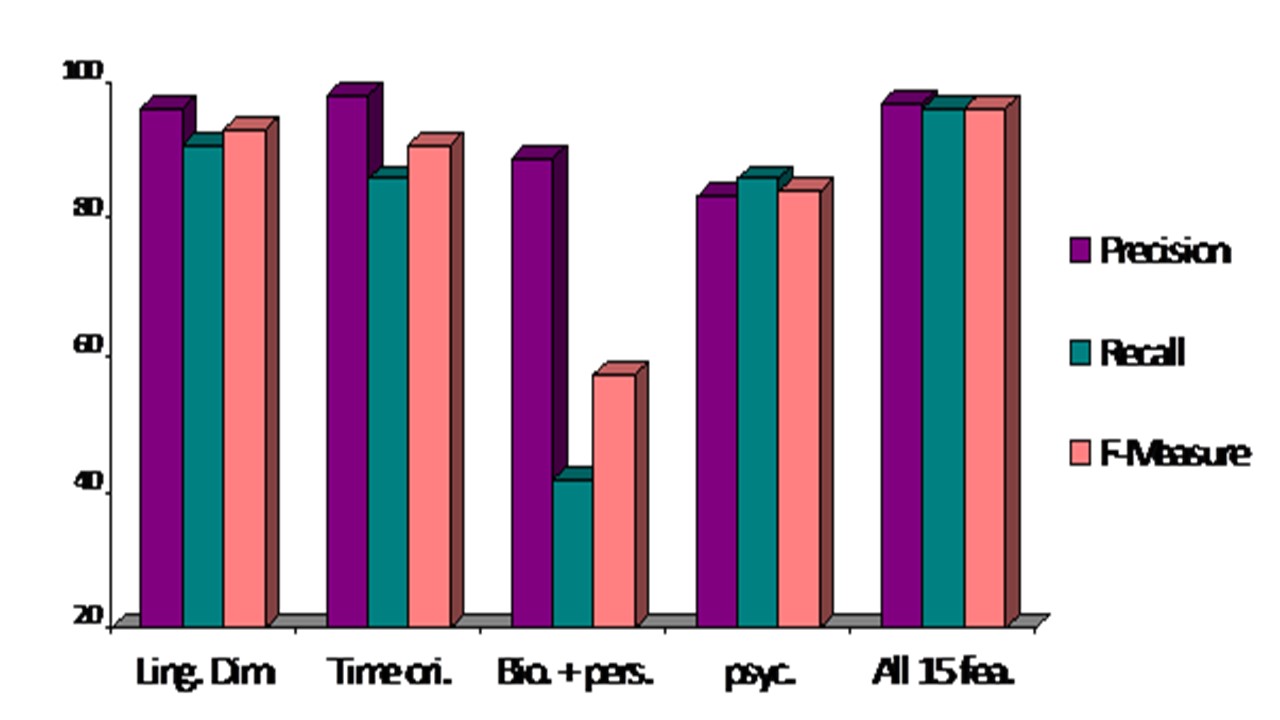}
	\caption{SVM classifier’s performance metrics on different feature sets}
	\label{fig:graph1}
\end{figure}

Table \ref{conf:liwc} and Fig. \ref{fig:graph1} show the various evaluation metrics of SVM classifier with the combination of various selected features. Higher the value of accuracy, the selected features are very good in prediction of classes of our model.

\begin{table}[]
	\centering
	\caption{Confusion Matrix of various combinations of LIWC features}
	\label{conf:liwc}
		\resizebox{\linewidth}{!}{
	\begin{tabular}{|l|l|l|l|l|}
		\hline
		\multicolumn{1}{|c|}{\textbf{SVM Classifier features set}}                                        & \multicolumn{1}{c|}{\textbf{Precision(\%)}} & \multicolumn{1}{c|}{\textbf{Recall(\%)}} & \multicolumn{1}{c|}{\textbf{F-Measure(\%)}} & \multicolumn{1}{c|}{\textbf{Accuracy(\%)}} \\ \hline
		\begin{tabular}[c]{@{}l@{}}Linguistic Dimensions\\ (3 features)\end{tabular}                      & 96                                          & 91                                       & 93                                          & 94                                         \\ \hline
		\begin{tabular}[c]{@{}l@{}}Time orientations \\ (3 features)\end{tabular}                         & 98                                          & 86                                       & 91                                          & 92                                         \\ \hline
		\begin{tabular}[c]{@{}l@{}}Biological Processes + \\ personal concern\\ (4 features)\end{tabular} & 89                                          & 42                                       & 57                                          & 68                                         \\ \hline
		\begin{tabular}[c]{@{}l@{}}Psychological Processes \\ (5 features)\end{tabular}                   & 83                                          & 86                                       & 84                                          & 84                                         \\ \hline
		\begin{tabular}[c]{@{}l@{}}Selected LIWC features \\ (15 features)\end{tabular}                   & 97                                          & 96                                       & 96                                          & 97                                         \\ \hline
	\end{tabular}
}
\end{table}

\begin{figure}
	\centering
	\includegraphics[width=1.0\linewidth]{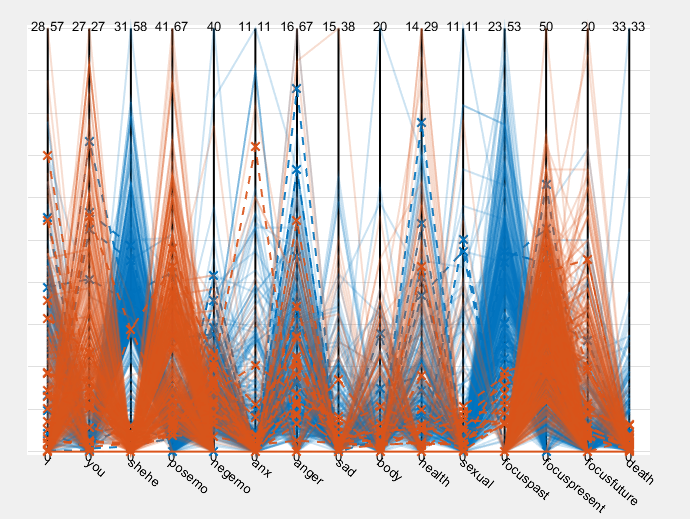}
	\caption{parallel coordinates plot for abuse and advice class}
	\label{fig:parallelplot}
\end{figure}

Parallel coordinates plot as shown in Fig. \ref{fig:parallelplot}, with all the selected features that separates the class value best. For example, ``posemo, focuspresent, you" are the features best classifying the class to be in ``advice/opinion", which is plotted in blue color. The orange color plot explains the class to be in ``abuse", with the selective features such as ``shehe, focuspast and death". We can understand that, when the victim or survivor posts about abusive experience, they use more past tense and health concern. `Shehe' notion also widely used to represent the abuser. Whereas, in the case of advice/opinion category, the linguistic style contain present tense and future tense, as it is more focused on future life and well-being.

\subsection{Term-Class Interaction model} \label{sec:predict1}

\begin{figure}
	\centering
	\includegraphics[width=1.0\linewidth]{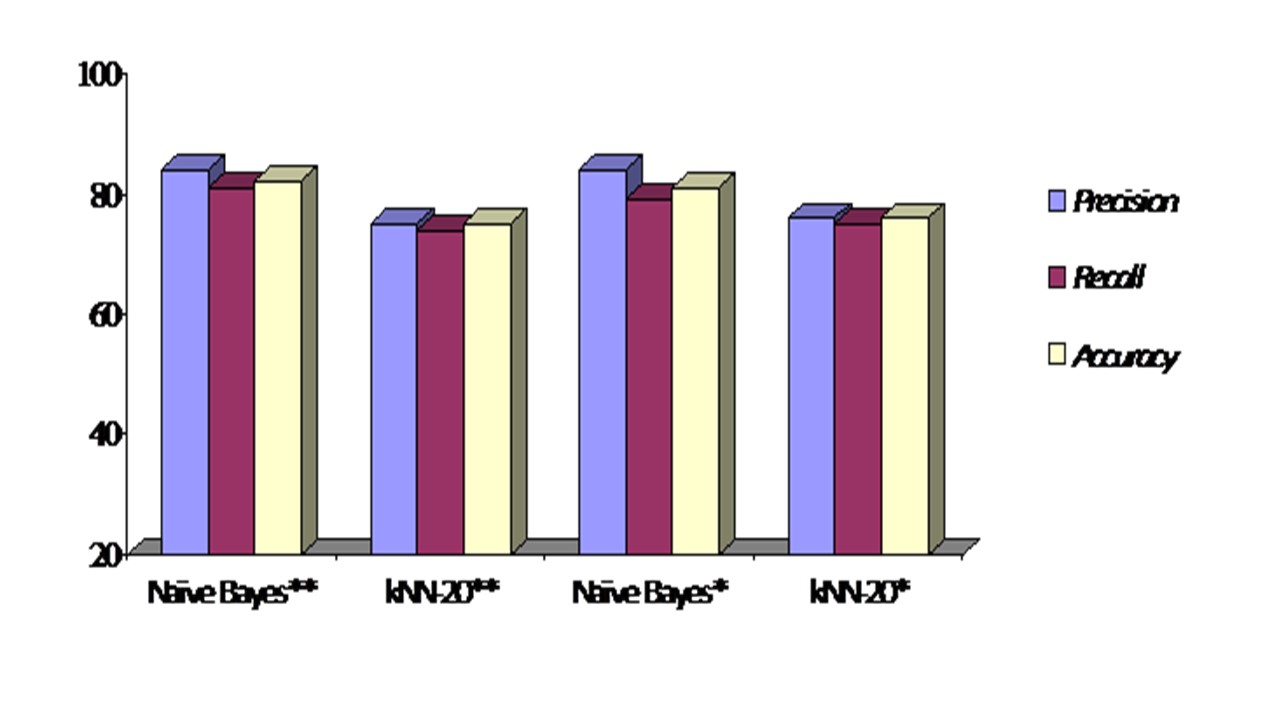}
	\caption{Precision, Recall and Classification accuracy of naive bayes and kNN (** 10 folds and * leave one out)}
	\label{fig:graph2}
\end{figure}

Our final model contains 300 features based on tf.idf vector and the total number of features is reduced to top 250 features based on the chi-squared value. Hence the feature space is significantly reduced. Finally standard Classifiers such as naive-bayes and k-NN are applied to classify the posts in the corresponding class and the results of the classifiers are compared with respect to standard evaluation metrics precision, recall and accuracy. Among the validation methods of leave one out and k folds cross validation, 10 folds cross validation is used in the final model for the better evaluation of the predictive accuracy. Our result in Fig. \ref{fig:graph2}. shows that classification accuracy of NB is 82\% with 10 folds cross validation, which outperforms  the classification accuracy of kNN.

The term-class interaction values using chi-squared test is shown in Table \ref{chi-squ-table}.  We can clearly see from Table \ref{chi-squ-table}, that terms, such as share, support, page, thank, are highly associated with class ``opinion/advice", whereas the terms, such as kill, husband, murder, leave, are more dependent with the class ``abuse".

\begin{table}[]
	\centering
	\caption{The selected terms with the Chi-squared value for predicted class}
	\label{chi-squ-table}
	\resizebox{\linewidth}{!}{
	\begin{tabular}{|l|l|l|l|l|l|}
		\hline
		\textbf{Term t} & \textbf{Chi-Squared value} & \textbf{Predicted class} & \textbf{Term t} & \textbf{Chi-Squared value} & \textbf{Predicted class} \\ \hline
		Leave           & 128.88                     & Abuse                    & Break           & 63.02                      & Abuse                    \\ \hline
		Kill            & 127.12                     & Abuse                    & Murder          & 61.5                       & Abuse                    \\ \hline
		Husband         & 98.63                      & Abuse                    & Share           & 21.09                      & Opinion                  \\ \hline
		Abuse           & 90.15                      & Abuse                    & Thank           & 20.99                      & Opinion                  \\ \hline
		Lose            & 75.01                      & Abuse                    & Support         & 16.62                      & Opinion                  \\ \hline
	\end{tabular}
}
\end{table}

\begin{figure}
	\centering
	\includegraphics[width=1.0\linewidth]{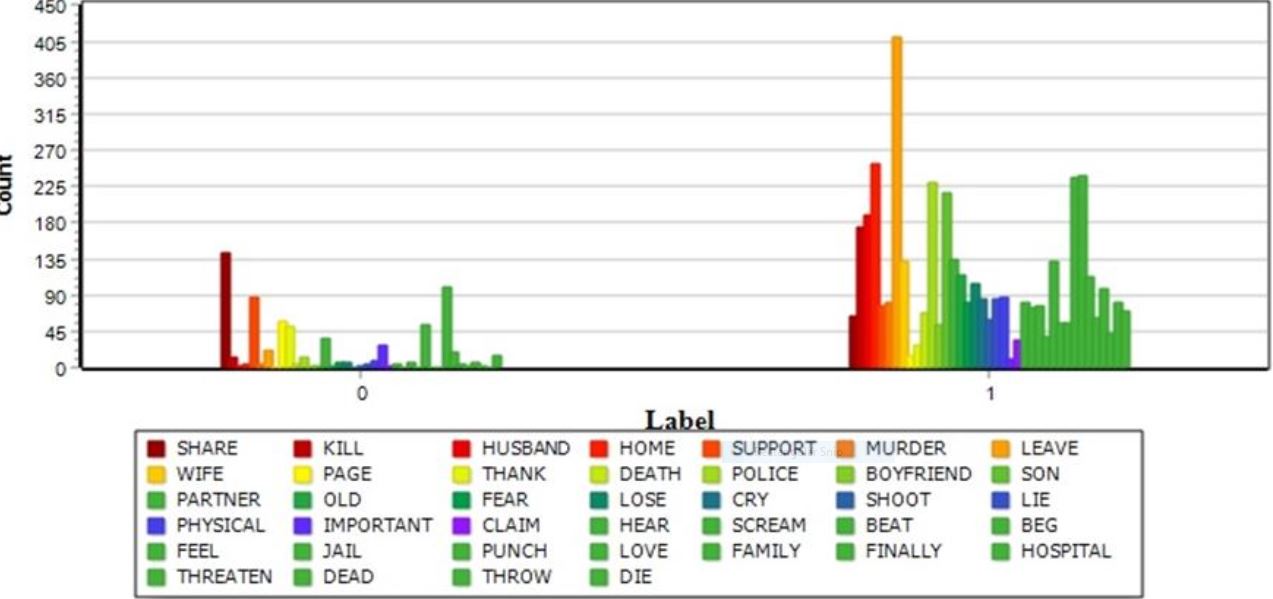}
	\caption{Visualization graph of the term-class interaction based feature selection method}
	\label{fig:word_result}
\end{figure}

Utilizing the property of $\chi^2$  statistic, it is inferred that higher the $\chi^2$ value of term t, indicates the higher likelihood of occurrence in the class c. Thus we use  $\chi^2$ metric to weight the context words in the tf.idf  model. The key aspect is that words with higher $\chi^2$ statistics tend to be keywords for class identification. Hence we applied chi-square statistical test to select the lexicon that particularly correlates to the specific class identification task for user posts. In this work, words that are likely to be valuable for the classification task are more heavily weighted based on  $\chi^2$ metric, and hence reducing the disturbance of the noise words which are not helpful comparatively to the later task. The Fig. \ref{fig:word_result} shows, each ter{m}'s probability in predicting the corresponding class. For example, the words, such as share and support belongs to class ``advice/opinion" (which is represented as 0). The words, such as kill, murder, leave, predicts the text to be in abuse class (represented as 1).

\section{Conclusion} \label{sec:concl}

The results of this study demonstrated that the linguistic dimensions and textual features discussed in the user posts have the potential to classify the text into appropriate class. The experimental results highlighted that psycholinguistic clues have strong indicative powers in the prediction of posts than textual features. By interpreting the use of proposed intent mining classification models, social support groups on Facebook can quickly identify DV victims via text posted on Facebook and appropriate support can be provided.
\bibliographystyle{IEEEtran}
\bibliography{refer}

\end{document}